\documentclass[12pt]{article}

\usepackage{graphics}
\usepackage{amssymb}

\newtheorem{lemma}{Lemma}
\newtheorem{teorema}{Theorem}

\def\min{{\rm min}}
\def\eps{ { \varepsilon } }

\def\phi{{\varphi}}
\def\equal{\buildrel {\rm def} \over {=} }
\def\={ \equal }
\newcommand{\vett}[1]{\mathbf{#1}} 
\newcommand{\Hf}{H_1} 
\newcommand{\rhof}{\rho_1}
\newcommand{\tq}{{\tilde{q}}}
\newcommand{\tV}{{\tilde{V}}}
\newcommand{\tf}{t_{inf}}
\newcommand{\dif}{\mathrm{d}}
\newcommand{\meanB}{\overline B} 

\title{Relaxation times for Hamiltonian systems}

\author{A.~Maiocchi\footnote{Universit\`a di Milano,
    Dipartimento di Matematica,
        Via Saldini 50, 20133 Milano, Italy.
       E-mail: \texttt{alberto.maiocchi@unimi.it}} \and A.~Carati\footnotemark[1]}

\begin{document}

\maketitle

\begin{abstract}
Usually, the relaxation times of a gas are estimated in the frame of  
the Boltzmann equation.
In this paper, instead, we deal with the relaxation problem in the 
frame  of the dynamical theory of Hamiltonian systems, 
in which the definition itself
of a relaxation time is an open question.
We introduce a lower bound for the relaxation time, and give a
general theorem for estimating it. 
Then we give an application to a concrete model of an
interacting gas, in which the
lower bound turns out to be of the order of magnitude of the
relaxation times observed in dilute gases.
\end{abstract}

\section{Introduction} 

The definition and the estimate of relaxation times are
problems of central interest when one attempts  describing macroscopic
systems through microscopic Hamiltonian models. 

In the case of gases, these problem are tackled, and solved, 
in the frame of the Boltzmann equation (see \cite{cercignani}). In
such a frame the existence 
of a relaxation time is somehow obvious, due to the irreversible character of
the equation, and the estimate is obtained in terms of the eigenvalues
of the linearized equation, about the equilibrium solution. On the
other end, Boltzmann equation refers to a reduced description, while
we want to tackle the problem considering the complete system.
This would require to estimate the time needed for an initial measure
in phase space to relax to an asymptotic one. This approach was
followed, for example, in \cite{gallavotti}, for a reversible
dissipative model, which should mimic a coupling of the system of
interest with two (or more) heat reservoirs.

In the present work, instead, we tackle the problem from the point of view of
the dynamical theory of Hamiltonian systems, for systems which are isolated. 
In this perspective, 
a partial answer to
the problem is given by Kubo's linear response theory
\cite{kubo}. Indeed, such a  theory enables one, at least in
principle, to compute in microscopic terms the macroscopic transport
 coefficients, and then,  via macroscopic equations, the
relaxation time. From our point of view, however, this answer is not completely
satisfactory, because it appeals to macroscopic irreversible
equations, which should preliminarily be deduced from the microscopic ones.

A related but different approach is followed here, whose main scheme can
be  sketched as follows.

From linear response theory we take the starting point, namely, the
idea of following the time evolution
of the probability distribution function in 
phase space (and not in the reduced $\mu$--space, as in the Boltzmann
equation), when a
perturbation $-hA(p,q)$ to the original Hamiltonian $H_0(p,q)$  is
introduced. Still following Kubo, we then choose to concentrate our
attention on a particular observable, namely the one conjugated to the
perturbing field, in the familiar sense in which pressure is conjugated
to volume and magnetization to the magnetic field. As an example,
later on in this paper we will deal with the simple case
in which the perturbing field is gravity, and the conjugate observable is
the height of the center of mass. Then, our
attention is addressed at defining and estimating the
relaxation time.

In the spirit of the Kubo approach it is natural to say that equilibrium
is attained when the time derivative of the distribution function is
negligible. The aim of the present paper is indeed to give a lower
bound $\tf$ to the relaxation
time, looking at the evolution not of the distribution function
itself, but of the time--derivative of the variable
conjugated to the perturbation $A$ in the Hamiltonian, which is
strictly related to the distribution function (see formula
(\ref{evoluzione_rho_1}) below). It is easily
seen that the time--derivative of the
conjugate variable is  the function $B\=[H_0,A]$, namely, the
time--derivative of $A$ with respect to the flow generated by the full
Hamiltonian $H$ (here, $[\cdot, \cdot ]$ denotes Poisson
bracket), so that this is the
quantity on which we will concentrate in  this paper.

Having chosen the relevant function, namely $B$, we make use of the
easily established properties (see later) that
its expectation vanishes at equilibrium, and that its time--derivative
is positive at the initial time. Thus a lower bound $\tf$ to the
relaxation time is provided by the time before which the
time--derivative of the expectation of $B$ is proven to be positive.

The problem is then that one should make use of suitable a priori
estimates on the dynamics, since an explicit integration of the
equations of motion is lacking. This can actually be implemented
following the main idea
introduced in paper \cite{carati}, which was concerned with 
Hamiltonian perturbation theory in the thermodynamic limit.
In such a paper, a procedure is given which,
for any $\mathcal{L}^2$ function $f$ of phase space with respect to
the Gibbs measure, allows
one to provide an upper bound to $\|U_tf-f\|_2$, by knowing an upper
bound to $\|[f,H]\|_2$. Here, $H$ is the Hamiltonian of the system,
and $U_t$ the
corresponing unitary evolution group. 

The estimate of the lower bound $\tf$ is provided by formula
(\ref{tempo_equilibrio}) of Theorem~\ref{teorema_rilassamento}, which
is stated and proved in Section~2. Such a proof is given for an ample
class of Hamiltonian systems, which are the ones considered in most 
rigorous works in Statistical Mechanics (see \cite{ruelle}).

In Section~3 the general theorem is applied to the case of a gas of
interacting point--particles enclosed in a cubic box, to which
the gravity force is added as a perturbation. To this aim, we give an
interesting estimate of the $s$--point correlation function for a gas
interacting through a stable and tempered two--body potential, which
is here obtained by extending some old results of Bogolyubov et
al. \cite{bogoljubov} and of O. Penrose \cite{penrose}. The lower bound to the
relaxation time thus found turns out to be comparable with the
typical  relaxation times observed in dilute gases. 

Some further comments are given in Section~4.

\section{General Theorem about Relaxation Times}

We consider an isolated Hamiltonian system, with phase space $\mathcal{M}$,
and with an invariant measure with respect to the unperturbed
Hamiltonian $H_0$. One could think that, in principle, one
has to take the microcanonical measure, but, in view of the ensemble
equivalence for large $N$ (see, for instance, \cite{minlos}), we will
take, instead, the Gibbs measure at inverse temperature $\beta$,
 i.e.,  the measure with density $\rho_0$ given by 
 $$
\rho_0=\frac{1}{Z_0}\exp\left(-\beta H_0\right)\ ,
$$ 
$Z_0$ being the partition function.
Suppose at time 0 a perturbation $-hA(p,q)$ is introduced, where $A$
is a given function on phase space and the parameter
 $h>0$ controls the size of the perturbation. So, 
at positive times the Hamiltonian  is $\Hf=H_0 -hA$. The corresponding
Gibbs density (at the same $\beta$) will be denoted by $\rho_1$. 
Our aim is to find a sensible lower bound for the relaxation time to
the final  equilibrium with respect to the full Hamiltonian $\Hf$.

To this end, along the scheme sketched in
the Introduction, following Kubo we 
consider the  observable $B$ defined by
$$
B=[A,H_0]\ ,
$$ 
i.e., the time derivative of the perturbation $A$ with respect to the
flow generated by the full Hamiltonian $\Hf$. We then consider the
probability density $\rho$, solution of Liouville's equation relative
to the total
Hamiltonian $\Hf$ with initial condition $\rho(0)=\rho_0$, and look at
the evolution of the  expectation of $B$, i.e., we
look at the quantity
$$
\meanB (t)= \int B\,\rho(t)\,\dif p\,\dif q \ .
$$
The quantity of interest actually will be its increment
$$
\Delta \meanB (t)\=\meanB (t)-\meanB (0)\ . 
$$
Writing $\rho$ in the form 
\begin{equation}\label{scomposizione_distribuzione}
\rho(t)\=\rho_0+\Delta\rho(t)\ ,
\end{equation}
 one has
\begin{equation}\label{integrale_DeltaB}
\Delta \meanB (t)= \int B\,\Delta\rho(t)\,\dif p\,\dif q \ .
\end{equation}
We will show that under the familiar conditions which 
entail  reversibility (namely, that both $H_0$ and $A$
are even in the momenta), the quantity $\Delta \meanB$ vanishes not
only (as it is obvious)  at
time zero, but also at equilibrium with respect to the full
Hamiltonian $\Hf$. This is due to the fact that 
 the expectations
of $B$ with respect to the Gibbs densities $\rho_0$ and $\rho_1$
corresponding to the Hamiltonians
$H_0$ and $\Hf$, both vanish by symmetry, because $\rho_0$ and
$\rho_1$ are  even in
  the momenta, whereas $B$ is odd. On the other hand, it turns out
 that $\Delta \meanB$ is initially an increasing function of
time, since its time--derivative is positive at time
0, as it will be shown later. Thus, the time--derivative of $\Delta
\meanB$ has to become negative at some time if
equilibrium with respect to the full Hamiltonian has to be
attained, and consequently   
a lower bound to the relaxation time is provided by
the time $\tf$ up to which   the time derivative of
$\Delta \meanB$ is guaranteed to be positive.

We thus  define  the lower bound $\tf$ by 
\begin{equation}\label{definizione}
\tf\=\sup\, t^*, 
\end {equation}
where $t^*$ is such that
\begin{equation}\label{definizione_bis}
\frac{\dif}{\dif t}\Delta\meanB (t)> 0\quad
\mbox{for all}\quad 0<t<t^*\ ,
\end{equation}
or $\tf=+\infty$ if
$$
\frac{\dif}{\dif t}\Delta\meanB (t)> 0\quad \forall t>0\ .
$$
Notice that our definition makes sense also for $h=0$, in which case
one has $\tf=0$, as can be seen by formula
(\ref{evoluzione_variazione}) below.

The problem is then to estimate the rate of growth   of $\Delta
\meanB$. Now, on the one hand, following Kubo we know that
$\Delta\meanB (t)$ is strictly related to the time--autocorrelation of $B$
(see (\ref{evoluzione_variazione}) below).
 On the other hand, we can  make use of the main result obtained in paper
\cite{carati}, in which it was shown how to estimate 
the time--autocorrelation of $B$ in terms of the Hamiltonian.
Indeed, from the main result of that paper one easily obtains the
following property: an  a priori estimate of the type
$$
\left\|\left[B,H_0\right]\right\|_0\le \eta \left\|B\right\|_0\ ,
$$
(with the norm defined below)
implies that the time--evolution of $\Delta \meanB$ is slow if $\eta$ is
small, or, more precisely, that the lower bound $\tf$ to the
relaxation time defined by
(\ref{definizione}), (\ref{definizione_bis}) is inversely proportional
to $\eta$.

Here,  $\|\cdot\|_0$ is the norm on $\mathcal{L}_0^2(\mathcal{M})$,
the Hilbert space of square
integrable complex functions on $\mathcal{M}$, with respect to $\rho_0$. 
We will also have to consider the Hilbert space $\mathcal{L}_1^2(\mathcal{M})$ 
of the square integrable complex functions with respect to $\rhof$. The
corresponding $\mathcal{L}^2$--norm will be denoted  by $\|\cdot\|_1$.

Under the rather natural condition (\ref{ipotesi1}) given below,
which ensures  the smallness of the 
``change''  of the Gibbs measure induced by the perturbation $-hA$, for a
large class of observables it can be proven that the two norms just introduced
are asymptotically equivalent as $h\to 0$. Indeed one has the
following lemma, whose proof is deferred to Appendix~A.
\begin{lemma}\label{lemma}
Assume  there exist $\delta>0$ such that
\begin{equation}\label{ipotesi1}
 \int_\mathcal{M} \dif p\,\dif q\, e^{\delta
     A}\rho_0<+\infty\quad\mbox{and}  \quad\displaystyle\int_\mathcal{M}
     \dif p\,\dif q\,
   e^{-\delta A} \rho_0<+\infty\ . 
\end{equation}
Then, for all  real functions $f$ on $\mathcal{M}$
satisfying at least one of the conditions
\begin{equation}\label{cond2}
\left\|f^2\right\|_0<+\infty\ ,\qquad\left\|f^2\right\|_1<+\infty\ ,
\end{equation}
one has
$$
\left\|f\right\|_1^2-\left\|f\right\|_0^2=o(1)\ ,\quad\mbox{as}\; h\to
0\ .
$$
\end{lemma}

We are now able to give an estimate for $\tf$ in terms of $\eta$,
which is provided  by the following  Theorem~1. It will be seen that some
technical hypotheses, namely those given in (\ref{ipotesi_tecniche}) below,
are required just in order that at least one of the conditions
(\ref{cond2}) of Lemma~\ref{lemma} is satisfied.
 
\begin{teorema}\label{teorema_rilassamento}
Let the unperturbed Hamiltonian $H_0(p,q)$ be even in the momenta and
bounded from below, and consider a perturbation $-h A(p,q)$, with $h>0$ and
$A$ even in the momenta. Suppose $A$ and $H_0$ are such that 
 hypothesis (\ref{ipotesi1}) of Lemma~1 is satified. With $B=[A,H_0]$,
suppose furthermore that the following  technical conditions are satisfied:

\begin{equation}\label{ipotesi_tecniche}
\left\|B^4\right\|_0<+\infty\ ;\quad
    \left\|\left[B,H_0\right]^2\right\|_0<+\infty\ ;\quad
    \left\|\left[B,A\right]^2\right\|_0<+ \infty\ .
\end{equation}
Then a lower bound to the relaxation time defined by
(\ref{definizione}), (\ref{definizione_bis}) is given by
\begin{equation}\label{tempo_equilibrio}
\tf\ge\frac{\sqrt{2}}{\eta}+o(1),\quad \mbox{as}\quad h\to 0\ ,
\end{equation}
where $\eta$ is such that
\begin{equation}\label{ipotesi2}
  \left\|[B,H_0]\right\|_0< \eta
   \left\| B\right\|_0\ .
\end{equation}
\end{teorema}

\textbf{Remark.} It may appear that some conditions are too
restrictive if the theorem has to be used in the
thermodynamic limit, but it turns out that such a difficulty can be
overcome. For example, $H_0$ was required to have a finite
lower bound, call it $D$; however, the result is found not to depend
on the value of $D$. So,
$D$ can grow with the number $N$ of degrees of freedom,
without affecting the validity of the theorem, provided $D$ is finite
for any finite $N$. A similar argument also applies to conditions 
(\ref{ipotesi1}) and (\ref{ipotesi_tecniche}), so the theorem holds
for any  system, however large it may
be. Then, in order to pass to the thermodynamic limit, it suffices to have
that $\eta$ tends uniformly to a finite limit as $N$ increases.
\smallskip

\textbf{Proof of Theorem~\ref{teorema_rilassamento}.} First of all, we
 notice  that  the time evolution  of the perturbation $\Delta \rho$ satisfies,
 by Liouville's equation, the differential equation
\begin{equation}\label{equazione_differenziale_rho}
\frac{\partial \Delta\rho}{\partial
  t}=\left[H_0-h A,\Delta\rho\right]-h\left[A,\rho_0\right]\ , 
\end{equation}
with $\Delta\rho(0)=0$ as initial condition. Such an equation admits a
unique solution in the Hilbert space
$\mathcal{L}^2_1( \mathcal{M})$.
Indeed, equation (\ref{equazione_differenziale_rho}) 
is a linear inhomogeneous
first--order differential equation in $\mathcal{L}^2_1( \mathcal{M})$
of the form
$$
\dot{x}=\hat{\mathcal{O}}x+f \ ,
$$
where the operator $\hat{\mathcal{O}}\=[H_1,\cdot]$ generates a
semigroup of unitary evolution transformations (see for example
\cite{koopman}). Thus, since the second term $h[A,\rho_0]$ at the r.h.s.
belongs to $\mathcal{L}^2_1(\mathcal{M})$, as will be shown below,
the solution is known to exist and be unique (see Theorem~3.3, page 104, of
  \cite{showalter}).
 Such a solution is given by a simple adaptation of the  variation of 
constants formula, namely by
\begin{equation}\label{evoluzione_rho_1}
\Delta\rho(\vett x,t)=\beta h\int_0^t ds\,B(\Phi^s \vett x)
\rho_0(\Phi^s \vett x)\ ,
\end{equation}
where $\vett x \= (p,q)$ denotes a point of phase space $\mathcal{M}$,
and $\Phi^t$  the flow generated by $H_1$.
Notice that, as the initial datum vanishes, one obviously has  
$\hat{\mathcal{O}}\, \Delta \rho(0)\in \mathcal{L}^2_1(\mathcal{M})$. 

We  show now that $[A,\rho_0]\in\mathcal{L}^2_1(\mathcal{M})$, too.
To this end, we first  notice that
\begin{equation}\label{esistenza_soluzione}
\left\|[A,\rho_0]\right\|_1=\left\|B\rho_0\right\|_1\le
\frac{e^{\beta D}}{Z_0}\left\|B\right\|_1\ ,
\end{equation} 
where $D\=\inf_{p,q}H_0$.
On the other hand, iterating the Schwarz inequality gives
$$
\left\|B\right\|_0\le\left(\left\|B^2\right\|_0\right)^\frac{1}{2}
\le\left(\left\|B^4\right\|_0\right)^\frac{1}{4}<+\infty\ ,
$$
in which  the first hypothesis of
(\ref{ipotesi_tecniche}) was used.\footnote{According  to the same reasoning,
  the square of the norm of a function will be bounded from above
  by the norm of the
  squared function.} By virtue of such an hypothesis,
 we can also apply Lemma~\ref{lemma} to $B$ and
observe that, for $h$ small enough, $\left\|B\right\|_1$ is
finite. Thus, by  (\ref{esistenza_soluzione}) it is proved that 
$[A,\rho_0]$ belongs to $\mathcal{L}^2_1(\mathcal{M})$.

We now look at the expectation $\meanB (t)$ and at its increment
$\Delta \meanB (t)$. By using (\ref{evoluzione_rho_1}) for $\Delta\rho$ in
(\ref{integrale_DeltaB}), one finds for $\Delta \meanB (t)$ the expression
\begin{equation}\label{evoluzione_variazione}
\Delta \meanB (t)=\beta h\int_\mathcal{M} \dif \vett x
\int_0^t ds\,B(\Phi^s \vett x)\, B(\vett x)
\rho_0(\Phi^s \vett x)\ .
\end{equation}
Using the shorthand $f(\vett x_t)=f(\Phi^{-t}\vett x)$, one has then:
$$
\frac{\dif}{\dif t}\Delta \meanB (t)=\beta h\int_\mathcal{M} \dif 
\vett x\,
B(\vett x_{-t})\,B(\vett x)\,\rho_0(\vett x_{-t})\ ,
$$
or equivalently (due to preservation of Lebesgue measure),
\begin{equation}\label{derivata_A_punto}
\frac{\dif}{\dif t}\Delta \meanB (t)
=\beta h\int_\mathcal{M} \dif 
\vett x\,
B(\vett x_t)\,B(\vett x)\,\rho_0(\vett x)\ .
\end{equation}

At this point we remark that the integral in (\ref{derivata_A_punto})
could be evaluated in a quite simple way, if there appeared $\rho_1$
in place of $\rho_0$. Indeed, due to the unitarity of the flow, for
any $f$ in $\mathcal{L}^2_1(\mathcal{M})$ one would have
\begin{equation}\label{fluttuazione}
\int_\mathcal{M} \dif \vett x\,
f(\vett x_t)\,f(\vett x)\,\rho_1(\vett
x)=\left\|f\right\|_1^2-\frac{1}{2}\left\|f(\vett x_t)-f(\vett x)\right\|_1^2\
,
\end{equation}
and thus, on account of hypothesis (\ref{ipotesi2}) of the theorem, the
thesis  would
follow by using Theorem~1 of \cite{carati} (see below).

The rest of
the proof is devoted to show that the error made by taking $ \rho_0$
in place of $\rho_1$ is negligible in the limit $h\to 0$. To this
end, we suitably rewrite (\ref{derivata_A_punto}) in the form
\begin{eqnarray}\label{minorazione_derivata_1}
\frac{\dif}{\dif t}\Delta \meanB (t)=\!\beta h \left[ \int_\mathcal{M}
  \!\!\!\dif\vett x\, B(\vett x_t)\,B(\vett x)\,\rho_1(\vett x)-\int_\mathcal{M}
  \!\!\!\dif\vett x\, B(\vett x_t)B(\vett x)\left(\rho_1(\vett x)-\rho_0(\vett x)\right)\right]\nonumber\\
\ge\! \beta h \left[\int_\mathcal{M}
 \!\!\!\dif \vett x\, B(\vett x_t)\,B(\vett x)\,\rho_1(\vett x)-\left|\int_\mathcal{M}
  \!\!\!\dif \vett x\,B(\vett x_t)B(\vett x)\left(\rho_1(\vett x)-\rho_0(\vett x)\right)\right|\right]\ .
\end{eqnarray}

First, we show that the second term at the r.h.s.
vanishes as $h\to 0$. Indeed, by  Schwarz's inequality we have
\begin{equation}\label{prima_maggiorazione}
\left|\int_\mathcal{M}\!\!\!\dif\vett x\, B(\vett x_t)B(\vett x)\left(\rho_1(\vett
x)-\rho_0(\vett x)\right)\right|
\le
\left[\int_\mathcal{M}\!\!\!\dif\vett x \left(B^2(\vett x_t)B^2(\vett
  x)\right)\rho_1(\vett x)\right]^\frac{1}{2}\sqrt{\tilde{\gamma}(h)}\ ,
\end{equation}
where we have defined
$$
\tilde{\gamma}(h)\=\int_\mathcal{M} \dif \vett
x\,\left(\frac{\rho_0(\vett x)}{\rho_1(\vett x)}-1
\right)^2\rho_1(\vett x)\ .
$$
This function coincides with the one defined by (\ref{definizione_gamma_tilde_h}) in Appendix~A. As
there shown, one has $\tilde{\gamma}(h)\to 0$ as $h\to 0$. We then make use of
(\ref{fluttuazione}), by replacing $B^2$ for $f$ and neglecting the
negative term, in order to find an upper
bound to the r.h.s. of (\ref{prima_maggiorazione}):  one has, in fact,
$$
\left[\int_\mathcal{M}\dif \vett x\, \left(B^2(\vett x_t)B^2(\vett
  x)\right)\rho_1(\vett x)\right]^\frac{1}{2}\le
\left\|B^2\right\|_1\ .
$$
In order to show that $\|B^2\|_1$ is finite, we use Lemma~\ref{lemma}, whose
hypotheses are satisfied
owing to the first inequality of (\ref{ipotesi_tecniche}). Thus, as 
$\tilde{\gamma}(h)\to 0$ for $h\to 0$, one gets
\begin{equation}\label{maggiorazione_secondo_addendo}
\left|\int_\mathcal{M}\dif\vett x\, B(\vett x_t)B(\vett x)\left(\rho_1(\vett
x)-\rho_0(\vett x)\right)\right|
=o(1)\quad\mbox{as}\quad h\to
0\ .
\end{equation}

We then come to the first term at the r.h.s. of
(\ref{minorazione_derivata_1}), which, using (\ref{fluttuazione}) again, can
be estimated as
\begin{equation}\label{fluttuazione_B}
  \int_\mathcal{M} \dif\vett x\, B(\vett x_t)\,B(\vett x)\,\rho_1(\vett x)
  =\left\|B\right\|_1^2-\frac{1}{2}\left\|B(\vett x_t)-B(\vett x)\right\|_1^2\ . 
\end{equation}
We now make use of Theorem~1 of \cite{carati}, which ensures that, if
$$
\left\|\left[B,\Hf\right]\right\|_1<\tilde{\eta}\left\|B\right\|_1$$
is satisfied, then one has
\begin{equation}\label{stima_variazione}
\left\|B(\vett x_t)-B(\vett x)\right\|_1< \tilde{\eta}
t\left\|B\right\|_1\ .
\end{equation}
Now, to give an estimate for $\tilde{\eta}$, we notice that the
following  inequalities hold as $h\to 0$:
\begin{eqnarray*}
\left\|\left[B,\Hf\right]\right\|_1 & \le & \left\|\left[B,H_0\right]\right\|_1+h\left\|\left[B,A\right]\right\|_1\\
& \le & \left(\left\|\left[B,H_0\right]\right\|_0^2+o(1)\right)^\frac{1}{2}\\
&
&+h\left(\left\|\left[B,A\right]\right\|_0^2+o(1)\right)^\frac{1}{2}\
.
\end{eqnarray*}
Here, in the first line the triangle inequality was used, while the
second line is a consequence of Lemma~\ref{lemma}, the hypotheses of
which are satisfied by virtue of the second and the third
inequalities in (\ref{ipotesi_tecniche}). Hence, by hypothesis
(\ref{ipotesi2}) we obtain
\begin{eqnarray}\label{variazione_A_punto_norma_finale}
\left\|\left[B,\Hf\right]\right\|_1 & \le &
\left\|\left[B,H_0\right]\right\|_0+o(1)\left\|B\right\|_1\nonumber\\
&< &
\eta\left\|B\right\|_0+o(1)\left\|B\right\|_1\nonumber\\ 
&\le
&\left(\eta+o(1)\right)\left\|B\right\|_1\ ,
\end{eqnarray}
so that $\tilde{\eta}=\eta+o(1)$.

Eventually, by replacing in (\ref{minorazione_derivata_1}) the
estimates (\ref{maggiorazione_secondo_addendo}) and
(\ref{fluttuazione_B}), one has
\begin{equation}\label{ultima_maggiorazione}
\frac{\dif}{\dif t}\Delta \meanB (t)\ge \beta
h\left(1-\frac{\eta^2t^2}{2}-o(1)-o(1)\cdot
t^2\right)\left\|B\right\|_1^2\ .
\end{equation}
Therefore, in the limit as $h\to 0$, the time derivative of
$\Delta \meanB(t)$ remains positive for
$$
t<\frac{\sqrt{2}}{\eta}+o(1)
$$
and this completes the proof.
\begin{flushright}Q.E.D.\end{flushright}

\section{A Gas in a Gravitational Field}

We study now a gas of $N$ interacting particles enclosed in a
tridimensional box of side $L$ and total volume $V=L^3$. Our aim is to
show that Theorem~1 holds if we take as a simple example of
perturbation the force of gravity, in which case
the conjugated variable will be the displacement on the vertical axis
of the center of mass of the system. 

For what concerns  the
interaction of the particles  with the walls, due to the form of the
conjugated variable  it  turns out that only the interaction with the
horizontal walls will matter. Thus we limit ourselves to choose a
particular form for the interaction potential with the horizontal
walls. The unperturbed Hamiltonian $H_0$ is then
\begin{equation}\label{definizione_Hamiltoniana_esempio}
 H_0\=\sum_{j=1}^N\frac{\vett
   p_j^2}{2}+U_N(\vett q_1,\ldots,\vett q_N)\ ,
\end{equation}
 where  $p_j^\alpha\in (-\infty,+\infty)$, $ q_j^\alpha\in
 (-L/2,L/2)$,  $\alpha=1,2,3$, $ q_j^3=z_j$, and $U_N$ denotes the
 potential energy of the system, which we take of the form
\begin{equation}\label{definizione_U_N}
  U_N\=\sum_{1\le i<j\le N}\Phi(\vett q_i-\vett
  q_j)+\sum_{i=1}^Nf(z_i)\ ,
\end{equation}
where $\Phi$ represents the mutual interaction
potential between the particles and $f$  the interaction
with the horizontal walls.

The choice of the possible forms of
$\Phi$ and $f$ is restricted by some technical conditions, which
guarantee the existence of a suitable upper bound to the
configuration integrals. In fact, the main difficulty which is
encountered in applying Theorem~\ref{teorema_rilassamento} to the present
case is the estimate of the distribution function for $s$
particles, often called the $s$--point correlation function.  We thus define
\begin{equation}\label{definizione_F_N_s}
  F^{(N)}_s(\vett q_1,\ldots,\vett q_s)\=V^s \int_V \dif^3 \vett
  q_{s+1}\ldots \int_V\dif^3 \vett q_N D_N(\vett q_1,\ldots,\vett
  q_N)\ ,
\end{equation}
where
\begin{equation}\label{definizione_D_N}
  D_N(\vett q_1,\ldots,\vett q_N)\=\frac{1}{Q_N}\exp\left[-\beta U_N(\vett
      q_1,\ldots,\vett q_N)\right]\ ,
\end{equation}
and
\begin{equation}\label{definizione_Q_N}
  Q_N\=\int_V\dif^3 \vett q_1\ldots \int_V\dif^3 \vett q_N
  \exp\left[-\beta U_N(\vett q_1,\ldots,\vett q_N)\right]\ . 
\end{equation}
For sufficiently low densities $\rho=N/V$, one can prove a useful
lemma which relates $F^{(N)}_s$ to
\begin{equation}\label{definizione_n_s}
  n_s(\vett q_1,\ldots,\vett q_s)\=\exp\left[-\beta U_s(\vett
      q_1,\ldots,\vett q_s)\right]\ ,
\end{equation}
under suitable conditions for the potentials $\Phi$ and
$f$. We consider $\Phi$ to be a stable and tempered potential in the
familiar sense (see, for example, \cite{ruelle}). Then, in
Appendix~\ref{dimostrazione_lemma_distribuzione} we prove the
following
\begin{lemma}\label{lemma_distribuzione}
Let $\beta$ be fixed. Let $\Phi$ be a stable potential, i.e. suppose
that exist $b>0$ such that
\begin{equation}\label{stabilità}
  \sum_{1\le i<j\le s}\beta\,\Phi(\vett q_i-\vett q_j)\ge -sb\ ;
\end{equation}
furthermore, assume that $\Phi$ is tempered and that $V$ is so large
that one has
\begin{equation}\label{definizione_I}
I\=\int_{\mathbb{R}^3}\left|e^{-\beta\Phi(\vett r)}-1\right|\dif^3
\vett r < \frac{e-2}{2(e+1)}V\ .
\end{equation}
Let $f$ be continuous on the open interval $(-L/2,L/2)$, nonnegative
and such that
\begin{equation}\label{definizione_L_tilde}
\tilde{L}\=\int_{-\frac{L}{2}}^\frac{L}{2}\, \dif z\, e^{-\beta f(z)} >
\frac{L}{2}\ .
\end{equation}
Then, for all densities $\rho$ satisfying
\begin{equation}\label{maggiorazione_rho_lemma}
\rho<\min\left[ \frac{1}{I}\left(\frac{\tilde{L}}{L}- \frac{1}{2}
  \right), \frac{1}{4e^{2b+1}I}\right] \ ,
\end{equation}
the following inequality holds.
\begin{equation}\label{risultato_lemma}
  F^{(N)}_s(\vett q_1,\ldots,\vett q_s)<
 \frac{2^s\sqrt{e}}{\sqrt{e}-1} \exp\left(\frac{sI_s}{2I}\right)  n_s(\vett
 q_1,\ldots, \vett q_s)\ ,
\end{equation}
where $I_1=I$ and, for $s\ge 2$,
\begin{equation}\label{definizione_I_s}
  I_s\=\int_{\mathbb{R}^3}\max \left\{\left(1 - e^{-\beta\Phi(\vett
    r)}\right) , \,e^{2sb} \left(1-e^{\beta\Phi(\vett
    r)}\right)\right\}\dif^3\vett r\ .
\end{equation}
\end{lemma}

In order to use such a lemma and have, at the same time, a
physically relevant model without complicating too much the
computations, we take $\Phi$ and $f$  equal to
the repulsive part of the Lennard--Jones potential, namely, given by
\begin{eqnarray}\label{definizione_potenziali}
\Phi(\vett x-\vett y) &\= & \frac{\eps}{\left\| \vett x - \vett
  y \right\|^{12}}\ ,\nonumber \\
f(z) &\=& \delta
   \left[\left(\frac{1}{z+\frac{L}{2}}\right)^{12}\!\!\!+\left(\frac{1}{z-\frac{L}{2}}\right)^{12}\right]\ ,
\end{eqnarray}
where $\eps$ and $\delta$ are positive parameters, chosen in a
convenient way (see hypothesis (\ref{scelta_parametri}) of Theorem~2).

According to the general scheme previously discussed, we add at time
$t=0$ a perturbation $-hA$, and for the observable $A$ we make the choice
\begin{equation}\label{definizione_perturbazione}
 A\=\sum_{j=1}^Nz_j\ .
\end{equation}

We then have
\begin{teorema}
Let $H_0$ and $A$ be given by
(\ref{definizione_Hamiltoniana_esempio}) and
(\ref{definizione_perturbazione}) respectively. Let the parameters $\eps$,
$\delta$ and $L$ in (\ref{definizione_potenziali}) be such that
\begin{equation}\label{scelta_parametri}
 (\beta\eps)^\frac{1}{12}<L/3 \quad \mbox{and}\quad
  (\beta\delta)^\frac{1}{12} < L/5\ ,
\end{equation}
where $\beta$ is  the inverse
temperature.  Then, for all densities $\rho$ which
satisfy
\begin{equation}\label{maggiorazione_rho}
  \rho<\frac{1.5\cdot 10^{-2}}{(\beta \eps)^\frac{1}{4}}\ ,
\end{equation}
one has the estimate $\tf\ge t_0+o(1)$, as $h\to 0$, with
\begin{equation}\label{tempo_equilibrio_esempio}
  t_0=\frac{1}{c}\left(\beta \delta\right)^\frac{1}{24} \sqrt{L\beta}
\end{equation}
and
\begin{equation}\label{costante_esempio}
c\=2\left(\frac{13e}{\sqrt{e}-1}\int_0^{+\infty}u^\frac{1}{12}e^{-u}
\dif u \right)^\frac{1}{2}
  \approx 14.5\ .
\end{equation}
\end{teorema}

\textbf{Remark 1.} The theorem is stated for a dimensionless
Hamiltonian. By inserting the proper dimensions, condition
  $(\beta\delta)^\frac{1}{12}<   L/5$  becomes
$(\beta\delta)^\frac{1}{12}\,\sigma<L/5$, where $\sigma$ is the
  charachteristic parameter for the range of the Lennard-Jones
  potential. This just expresses
the requirement that the interactions with the walls have a range
which is negligible with respect to $L$. Notice that we inserted the
factor $1/5$ at the r.h.s. just in order to fix a numerical value for
$c$, but any other reasonable choice would not affect the result. A
similar reasoning holds for the other requirement in
(\ref{scelta_parametri}). Notice, however, that the value of $t_0$
does not depend on the interaction potential\footnote{This is true for every potential $\Phi$
for which Lemma~\ref{lemma_distribuzione} can be applied (see the
discussion concerning (\ref{derivata_esempio_2}) and
(\ref{integrale_parti}) in the proof).} $\Phi$, which affects
only the density up to which the result is valid, through formula
(\ref{maggiorazione_rho}).

\textbf{Remark 2.} The time $t_0$, once the correct dimensional
constants have  been introduced, becomes
$$
t_0=\frac{1}{c}(\beta \delta)^\frac{1}{24}\sqrt{\beta m
  L\sigma}\ ,
$$
where $m$ is the mass (of molecular order) of each particle. Therefore, for
macroscopic systems in which $L$ is of the order of magnitude of $1\,
m$, while $\sigma\approx 10^{-10}\,m$, at room
temperature one gets $t_0\approx 10^{-8} s$, a value which is of the 
order of magnitude of the  typical relaxation times measured in
gases. In the same way, condition (\ref{maggiorazione_rho}) is proved
to hold for ordinary densities, namely, of the order of magnitude of
$10^{24}\, m^{-3}$.  Further comments will be given in the next section.

\smallskip

\textbf{Proof.} The proof consists in showing that the hypotheses of
Theorem~\ref{teorema_rilassamento} hold, with
$$
\eta =
  \frac{\sqrt{2}c}{\left(\beta \delta\right)^\frac{1}{24}
    \sqrt{L\beta}}\ .
$$

In the first place, (\ref{ipotesi1}) is satisfied for any $\delta$, because
it involves integrals of continuous functions over a compact, and  the
integral over the $\vett p$ coordinates is equal to 1.

As far as the other hypotheses are concerned, we first notice that in
the present case one has
$$
B=\sum_{j=1}^N p_j^z\ .
$$
The integral over the momenta of any power of $B$ can be easily
turned into a combination of terms of the form
$$
\int_{-\infty}^{+\infty}\dif p\,p^n\exp\left(\frac{-\beta p^2}{2}\right)\ ,
$$
which are finite for any $n$, thus proving the first of
(\ref{ipotesi_tecniche}). In particular, one has
\begin{equation}\label{derivata_esempio_1}
\left\|B\right\|_0=\sqrt\frac{2N}{\beta}\ .
\end{equation}
Clearly, one also has
$$
\left[B,A\right]=N\ ,
$$
and hence the third of (\ref{ipotesi_tecniche}) holds. We then
compute $\|[B,H_0]\|_0$. One has
\begin{equation}\label{derivata_esempio_2}
\left[B,H_0\right]= -\sum_{j=1}^N f'(z_j)\ ,
\end{equation}
with
$$
f'(z_j)=-12\delta\sum_{j=1}^N
\left[\left(\frac{1}{z_j+\frac{L}{2}}\right)^{13}+\left(\frac{1}{z_j-\frac{L}{2}}\right)^{13}\right]\
,
$$
since the contribution due to the pair interaction is the
$z$--component of the sum of all the internal forces, and consequently
it vanishes. 
Function (\ref{derivata_esempio_2}), and its
powers too, are actually singular at some point
in phase space, but they diverge there as a power,
while the density $\rho_0$  vanishes as an exponential, making the norm finite.
So the second of (\ref{ipotesi_tecniche}) is proved, as well.
  
There finally remains the task of providing an estimate for the
quantity $\eta$, which is an upper bound for the ratio
$\|[B,H_0\|_0/\|B\|_0$. To this end, from (\ref{derivata_esempio_2})
we get 
$$
[B,H_0]^2=\sum_{j,l=1}^Nf'(z_j)f'(z_l)\ ,
$$
and we have to estimate its $\rho_0$ norm.

It is convenient to
integrate by parts: we use the equality
\begin{eqnarray}\label{integrale_parti}
f'(z_j) f'(z_l) e^{-\beta U_N} &= & \frac{1}{\beta^2} \frac{\dif}{\dif z_l}
\frac{\dif}{\dif z_j}e^{-\beta U_N} - \frac{1}{\beta} \frac{\dif}{\dif
  z_l}\left( F_j^z e^{-\beta U_N}\right) -  F_l^z f'(z_j)e^{-\beta U_N}
  \nonumber \\
& &+\frac{\delta_{jl}}{\beta} f''(z_j) e^{-\beta U_N}\ ,
\end{eqnarray}
where $\delta_{jl}$ is the Kr\"{o}necker delta and
 by $F_j^z$ we denote the $z$ component of the force exerted on
the $j$--th particle by the other particles, i. e.
$$
F_j^z\= -\frac{\dif}{\dif z_j} \sum_{1\le i<k \le N} \Phi(\vett
q_i-\vett q_k)\ .
$$

We observe that the first term at the r.h.s. of
(\ref{integrale_parti}) has
a vanishing inegral, being the derivative of a function which vanishes
at the boundaries. Moreover, for what concerns the second term, we
point out that the quantity $\sum_j F_j^z$  vanishes, being
the $z$ component of the sum of all the internal forces. The same
remark is in order for the sum over
$l$ of the terms in the third place. The only term left is thus the
fourth one, which we write in the form $\delta_{jl}$ times the function
\begin{equation}\label{fine_integrale_parti}
\frac{1}{\beta}f''(z_j) e^{-\beta U_N}=\frac{156\,\delta}{\beta}
\left[\left(\frac{1}{z_j+\frac{L}{2}}\right)^{14}+\left(\frac{1}{z_j-
    \frac{L}{2}}\right)^{14}\right] e^{-\beta U_N}\ .
\end{equation}
Therefore, we have to compute $N$ identical integrals of the latter quantity,
depending only on one coordinate.

We remark here that conditions (\ref{scelta_parametri}),
(\ref{maggiorazione_rho}) are sufficient to ensure that
Lemma~\ref{lemma_distribuzione} can be used. In fact, our $\Phi$ is
certainly stable, tempered and nonnegative, while $f$ is continuous on
$(-L/2,L/2)$; furthermore, it is easy
to verify the remaining hypotheses, on account of a simple integration for
(\ref{definizione_I}) and of numerical computations of the integrals
appearing in
(\ref{definizione_L_tilde}) and (\ref{maggiorazione_rho_lemma}).

Making use of Lemma~{\ref{lemma_distribuzione}} for $F^{(N)}_1$ one
finds that the modulus of the integral of
(\ref{fine_integrale_parti}), which is independent of $j$, is
smaller than
\begin{equation}\label{integrale_esempio_1}
\frac{312\,e\,\delta}{(\sqrt{e}-1)\beta
  L}\int_{-\frac{L}{2}}^\frac{L}{2}\dif z\left[\left(\frac{1}{z +
    \frac{L}{2}} \right)^{14}+\left(\frac{1}{z-\frac{L}{2}}\right)^{14}
  \right]e^{-\beta f(z)}\ .
\end{equation}
The two terms in the integral (\ref{integrale_esempio_1}) are
identical, due to symmetry. Each of them is bounded from above on
account of the inequality
\begin{eqnarray*}
\int_{-\frac{L}{2}}^\frac{L}{2}\dif z\left(\frac{1}{z+\frac{L}{2}}\right)^{14}
e^{-\beta f(z)}& \le &\int_0^L\dif z\frac{1}{z^{14}}\exp\left(-\frac{\beta
  \delta}{z^{12}}\right)\\
&\le&\frac{1}{12(\beta\delta)^\frac{13}{12}}\int_0^{+\infty}u^\frac{
  1}{12} e^{-u}\dif u\ .
\end{eqnarray*}

In conclusion, we can infer that
\begin{equation}\label{primo_contributo}
  \left\|\left[B,H_0\right]\right\|_0^2 < \frac{N c^2}{\beta^\frac{25}{12}
    \delta^\frac{1}{12}}\ ,
\end{equation}
and a quick comparison with (\ref{derivata_esempio_1}) shows that one has
\begin{equation}\label{derivata_esempio}
  \left\|\left[B,H_0\right]\right\|_0< 
   \frac{\sqrt{2}c}{\left(\beta \delta\right)^\frac{1}{24} \sqrt{L\beta}} 
  \left\|B\right\|_0\ .
\end{equation}

This relation, on account of  Theorem~\ref{teorema_rilassamento},
leads to formula (\ref{tempo_equilibrio_esempio}).
\begin{flushright}Q.E.D.\end{flushright}

\section{Conclusions}
We have provided by Theorem~\ref{teorema_rilassamento} a lower bound
to the relaxation times in Hamiltonian systems, and shown that in the case
of an interacting gas enclosed in a box such an estimate is of the order of
magnitude of the  typical relaxation times measured in
gases.

This fact seems to indicate that the interactions
with the walls, which we have considered in the present work, might
have sensible effects even when one is interested
in investigating relaxations of observables related to internal
interactions.

We now add some comments concerning possible further developments.

The first point concerns the hypothesis made in
Theorem~\ref{teorema_rilassamento}, that $H_0$ and $A$
are even in the momenta, which actually is not at all essential. Indeed, if
such an hypothesis is not satisfied, it suffices to define $\tf$ in a
different way, namely, as the time up to which the
time--derivative of $\Delta \meanB$ remains larger than, for example,
$\frac{1}{2}\frac{\dif} {\dif t}\Delta \meanB(0)$. This way, by
inequality (\ref{ultima_maggiorazione}) one could prove
Theorem~\ref{teorema_rilassamento}, except for setting
$1/ \eta$ in place of $\sqrt{2}/ \eta$ in
(\ref{tempo_equilibrio}). We decided to deal with the case of
reversible Hamiltonians just because it is a very important one;
furthermore, in such a case the relaxation time can be defined with no
reference to arbitrary features, as the factor $1/2$ introduced above.

As a more interesting fact, we are confident that our line of reasoning
may be extended to the case of perturbations of a finite size
$h$, because this would just entail to consider the norms in
$\mathcal{L}^2_1(\mathcal{M})$ rather than in
$\mathcal{L}^2_0(\mathcal{M})$.  Indeed, if  we substitute $[B,H_1]$ for
$[B,H_0]$ in hypothesis (\ref{ipotesi2}) and use there the norm
\mbox{$\|\cdot\|_1$} instead of $\|\cdot\|_0$, we can directly set
$\tilde{\eta}=\eta$ in (\ref{stima_variazione}) and there is no need of
deducing  (\ref{variazione_A_punto_norma_finale}), so that the second
and  third  conditions in (\ref{ipotesi_tecniche}) are no more
required. Moreover, the first
condition in (\ref{ipotesi_tecniche}) can be replaced by  the condition that
$\|B^2\|_1$ is finite, which makes trivial the proof that $[A,H_0]$ is in
$\mathcal{L}^2_1(\mathcal{M})$.

The really open problem that remains in order to implement an
extension to the case of finite $h$, at least
for macroscopic systems, is the
estimate of the difference of the two norms in
(\ref{prima_maggiorazione}). The estimate which appears in such a formula
has the  serious flaw of increasing exponentially
with the number $N$ of particles. This occurs because the upper
bound provided there, which is an immediate consequence of the Schwarz
inequality, is valid for all functions in $\mathcal{L}^2$. A way
to improve such an estimate would be to restrict oneself to perturbations
having some suitable characteristic features.

In particular, the work \cite{lanford} of Lanford seems to suggest a
good starting point. There it is pointed out that the only observables of
interest in describing a macroscopic system are the ones he calls
finite range observables,\footnote{As a matter of fact, he explains
that this definition is chosen to make things simpler and is too
restrictive.  He gives also
a reference to Ruelle's book \cite{ruelle} in which it is shown how to
deal  with a broader class of observables, which represents the 
class of real  interest.} namely, observables which are sums of 
terms depending only
on the position of a finite number of particles. The difference of the
two norms in question can then be evaluated for each term and this
should lead to an
estimate which doesn't increase too much with $N$. We think that, if one
limits oneself to considering a smaller class of functions, there is a
good chance that the problem of the number of degrees of freedom is
overcome, and that some results are obtained
also for the case of perturbations of finite size.
These interesting investigations are left for possible future works.
\vspace{.8em}\\
{\large\textbf{Acknowledgements.}}
We thank very much Professor C.~Cercignani and L.~Galgani for useful
comments and discussions.

\appendix

\section{Proof of Lemma~\ref{lemma}}
We take as starting point the obvious equality
$$
\int_\mathcal{M} \dif p\,\dif q\, f^2\, \rho_1=\int_\mathcal{M} dp\,dq\,
f^2\,\rho_0+ \int_\mathcal{M} \dif p\,\dif q\,f^2\,(\rho_1-\rho_0)
$$
which, by the Schwarz inequality, gives 
\begin{equation}\label{minorazione_differenza_norme}
\left|\int_\mathcal{M}
\dif p\,\dif q\,f^2\,(\rho_1-\rho_0)\right|\le\left(\int_\mathcal{M}
\dif p\,\dif q\,f^4\, \rho_0\right)^\frac{1}{2}\sqrt{\gamma(h)}\ ,
\end{equation}
with
$$
\gamma(h)\= \int_\mathcal{M} \dif p\,\dif
q\,\left(\frac{\rho_1}{\rho_0}-1\right)^2\rho_0\ .
$$
One can also write
\begin{equation}\label{definizione_gamma_h}
\gamma(h)
=\frac{\int_\mathcal{M} \dif p\,\dif q\, e^{2h\beta A}\rho_0}{
  \left(\int_\mathcal{M} \dif p\,\dif q\, e^{h\beta
    A}\rho_0\right)^2}-1\ ,
\end{equation}
as is seen by expanding the square and using the fact
that $\rho_0$ and $\rho_1$ are the densities of the Gibbs
measures corresponding to $H_0$ and $H_1$, respectively. It is also of
interest to provide an upper bound to the l.h.s. of
(\ref{minorazione_differenza_norme}) in terms of $\rho_1$ rather than
of $\rho_0$. Indeed, one has
$$
\left|\int_\mathcal{M}
\dif p\,\dif q\,f^2\,(\rho_1-\rho_0)\right|\le\left(\int_\mathcal{M}
\dif p\,\dif q\,f^4\, \rho_1\right)^\frac{1}{2}\sqrt{\tilde{\gamma}(h)}\ ,
$$
where, in a way similar to (\ref{definizione_gamma_h}), one gets
\begin{eqnarray}\label{definizione_gamma_tilde_h}
\tilde{\gamma}(h)&\=&\int_\mathcal{M} \dif p\,\dif q\,\left(\frac{\rho_0}{\rho_1}-1\right)^2\rho_1\nonumber\\
&=&\left(\int_\mathcal{M} \dif p\,\dif q\, e^{h\beta
  A}\rho_0\right) \left(\int_\mathcal{M} \dif p\,\dif q\, e^{-h\beta
  A}\rho_0\right)-1\nonumber .
\end{eqnarray}

Now, we observe that the functions $\gamma(h)$ and $\tilde{\gamma}(h)$ can take
arbitrarily small values as $h$ goes to 0, if (\ref{ipotesi1}) is
satisfied.  Indeed,
by their definitions, they are always nonnegative quantities. Thus,
in order to show, for example, that $\gamma(h)<\varepsilon$  for any
fixed positive $\varepsilon$, it
will suffice that one has
$$
\frac{\int_\mathcal{M} \dif p\,\dif q\, e^{2h\beta
    A}\rho_0}{ \left(\int_\mathcal{M} \dif p\,\dif q\, e^{h\beta
    A}\rho_0\right)^2}<1+\varepsilon\ .
$$
To this end, let us note that
$$
\frac{1}{\int_\mathcal{M} \dif p\,\dif q\, e^{h\beta A}\rho_0}\le
\int_\mathcal{M} \dif p\,\dif q\,e^{-h\beta A}\ , 
$$
according to the Schwarz inequality. We combine this estimate with the
H\"{o}lder  inequality, on
whose account, if $h<\frac{\delta}{c \beta}$ and condition (\ref{ipotesi1}) holds, one has
\begin{eqnarray}\label{Hoelder}
  \int_\mathcal{M} \dif p\,\dif q\, e^{\pm c h \beta A}\rho_0
 & \le & \left(\int_\mathcal{M} \dif p\,\dif q\,e^{\pm\delta
 A}\rho_0\right)^\frac{c h \beta}{\delta} \left(\int_\mathcal{M} \dif
 p\,\dif q\,\rho_0\right)^{(1-\frac{c h \beta}{\delta})}\nonumber\\
 & \le& K^\frac{c h \beta}{\delta}\ ,
\end{eqnarray}
for some $K>0$, and we eventually obtain
\begin{eqnarray*}
\frac{\int_\mathcal{M} \dif p\,\dif q\, e^{2h\beta
    A}\rho_0}{ \left(\int_\mathcal{M} \dif p\,\dif q\,
    e^{h\beta A}\rho_0\right)^2}& \le &\left(\int_\mathcal{M} \dif
    p\,\dif q\, e^{2h\beta A}\rho_0\right)
    \left(\int_\mathcal{M} \dif p\,\dif q\, e^{-h\beta A}\rho_0\right)^2\\
&\le & K^\frac{4 h\beta}{\delta} \to 1\quad\mbox{as}\quad h\to 0\ .
\end{eqnarray*}
The immediate consequence is that, if we take
$h<\min\left(\frac{\delta}{2\beta},\frac{\delta \log
  (1+\varepsilon)}{4\,\beta \log K}\right)$,  then $\gamma(h)$ is
less than $\varepsilon$. An analogous argument, still based on inequality
(\ref{Hoelder}), ensures that $\tilde{\gamma}(h)$ takes arbitrarily small
values as $h$ goes to 0, too. Hence, the difference between
$\left\|f\right\|_0^2$ and $\left\|f\right\|_1^2$ vanishes with $h$,
provided $f^2$ belongs to $\mathcal{L}^2_0(\mathcal{M})$ or
$\mathcal{L}^2_1(\mathcal{M})$. \begin{flushright}Q.E.D.\end{flushright}

\section{Proof of Lemma~\ref{lemma_distribuzione}}\label{dimostrazione_lemma_distribuzione}

The lemma is proved by using the results that were obtained in
\cite{bogoljubov} in deducing the Mayer--Montroll equation.

A relevant difference from the classical works in this field is that
here we have
to deal with an external field, too. In order to connect the
computations with the ones used in the absence of such a field,
we change the coordinates from $\vett q_j$ to $\vett
\tq_j$, according to
\begin{equation}\label{cambiamento coordinate}
\tilde{q}^1_j  =  q^1_j\ , \quad
\tilde{q}^2_j = q^2_j\ ,\quad
\tilde{z}_j = \int_{-L/2}^{z_j}\dif x\, e^{-\beta f(x)}
\end{equation}
This change of coordinates is well defined, because $f$ is continuous
and the following inequality holds:
$$
\frac{\dif \tilde{z}_j}{\dif z_j}>0\qquad \forall z_j\in
\left(-\frac{L}{2}, \frac{L}{2}\right)\ .
$$
The volume differential, thus, changes in accordance with
$$
e^{-\beta f(z_j)}\dif^3 \vett q_j=\dif^3 \vett \tq_j\ ,
$$
making the external field disappear in the integration; we denote by
$\tV$ the domain of integration which takes the place of
$V$. We must also notice that the pair potential
$\Phi(\vett q_i-\vett q_j)$ is replaced by
$$
\tilde{\Phi}(\vett \tq_i, \vett
\tq_j) \=  \Phi(\tilde{q}^1_i-\tilde{q}^1_j,
\tilde{q}^2_i-\tilde{q}^2_j, z_i(\tilde{z}_i)-z_j(\tilde{z_j}))\ ,
$$
which is a function of the coordinates of both particles, and not only
of their difference, as was the case for $\Phi$, but it is still
symmetric under the exchange of $\vett \tq_i$ and $\vett \tq_j$.

Furthermore, we point out here that we must consider a different
distribution function in the modified phase space.  Such a function,
which we call $\tilde{D}_N$, is chosen by asking that it preserves the
volume element under the change of coordinates from $\vett q_i$ to
$\vett \tq_i$, i. e. that it fulfills the requirement
$$
\tilde{D}_N(\vett \tq_1,\ldots,
\vett \tq_N)\,\dif^3\vett \tq_1\ldots\dif^3\vett \tq_N =
 D_N (\vett q_1,\ldots,\vett q_N)
\,\dif^3\vett q_1\ldots\dif^3\vett q_N\ ,
$$
if $\vett \tq_i=\vett \tq_i(\vett q_i)$ for $i=1,\ldots ,N$.
It is apparent that one has
$$
\tilde{D}_N(\vett \tq_1,\ldots,\vett
\tq_N)=\frac{1}{Q_N}\exp\left[-\beta \sum_{1\le i<j\le N}
  \tilde{\Phi}(\vett \tq_i,\vett \tq_j)\right]\ .
$$
Accordingly, we will study the modified $s$--point correlation
$\tilde{F}^{(N)}_s$, defined in a way similar to
(\ref{definizione_F_N_s}), replacing $D_N$ with $\tilde{D}_N$ and
$\vett q$ with $\vett \tq$. The relation with the function
$F^{(N)}_s$, which is the one we want to bound from above, would be
\begin{equation}\label{F_tilde_N_s}
F^{(N)}_s(\vett q_1,\ldots,\vett q_s)=\tilde{F}^{(N)}_s(\vett
\tq_1(\vett q_1),\ldots,\vett \tq_s(\vett q_s))\,\exp\left(-\beta
\sum_{i=1}^s f(z_i)\right)\ .
\end{equation}

We can now repeat the deduction of the Mayer--Montroll equations in
these new coordinates, by writing
\begin{eqnarray}\label{mayer_montroll}
\tilde{F}^{(N)}_s(\vett \tq_1,\ldots,\vett \tq_s) & = & \frac{V^s
  Q_{N-s}}{ Q_N}\left[\int_\tV\dif^3 \vett
\tq^*_1 \ldots \int_\tV \dif^3 \vett \tq^*_{N-s}\,
\tilde{D}_{N-s}(\vett \tq^*_1,
\ldots,\vett \tq^*_{N-s})\right.\times\nonumber\\
& &\times\left.\prod_{i=1}^{N-s}\left(\tilde{f}_s(\vett \tq_1,\ldots,\vett
\tq_s;\vett \tq^*_i)+1\right)\right]\tilde{n}_s(\vett
\tq_1,\ldots,\vett \tq_s)\ ,
\end{eqnarray}
with
\begin{eqnarray*}
\tilde{f}_s(\vett \tq_1,\ldots,\vett
\tq_s;\vett {\tilde{y}}) &\=& \prod_{i=1}^s\exp\left[-\beta
  \tilde{\Phi}(\vett \tq_i,\vett {\tilde{y}})\right] -1\ ,\\
\tilde{n}_s(\vett \tq_1,\ldots,\vett
\tq_s) &\=&
\exp\left[-\beta\sum_{1\le i<j\le s}
  \tilde{\Phi}(\vett \tq_i,\vett \tq_j)\right]\ .
\end{eqnarray*}
We come then to the problem of finding an upper bound for the fraction
\mbox{$Q_{N-s}/Q_N$}  and for the
term in square brackets in (\ref{mayer_montroll}).

As far as the fraction is concerned, it is shown\footnote{See formulas
  (3.20) and (3.21) in \cite{bogoljubov}.} in
\cite{bogoljubov} that
\begin{eqnarray*}
  \frac{ Q_M}{Q_{M-1}} &\ge& \frac{1}{Q_{M-1}}\int_\tV \dif^3\vett \tq_1\ldots
  \int_\tV \dif^3\vett \tq_{M-1}\exp\left(-\beta \sum_{1\le i<j\le
      M-1} \tilde{\Phi}(\vett \tq_i,\vett \tq_j)\right)\times\\
& &\times \left[\int_\tV \dif^3 \vett \tq\left(1-\sum_{l=1}^{M-1}\left| 
  e^{-\beta\tilde{\Phi}(\vett \tq,\vett
    \tq_l)}-1\right|\right)\right]\ .
\end{eqnarray*}
Integrating over $\vett \tq$, we obtain that the term in square
brackets at the r.h.s. of this inequality is bounded from below by
$$
V\frac{\tilde{L}}{L}-(M-1)\sup_{\vett \tq_l\in \tV}\int_\tV \dif^3 \vett \tq\left| 
  e^{-\beta\tilde{\Phi}(\vett \tq,\vett \tq_l)}-1\right| \ge
  V\frac{\tilde{L}}{L}-(M-1)I\ ,
$$ 
with $\tilde{L}$ defined by (\ref{definizione_L_tilde}) in the
statement of Lemma~\ref{lemma_distribuzione}.
Therefore, if hypothesis (\ref{maggiorazione_rho_lemma}) holds, one has
$$
\frac{V Q_{M-1}}{Q_M}\le\frac{1}{\tilde{L}/L-\rho I}\le 2\ .
$$
Thus, we can write
\begin{equation}\label{maggiorazione_frazione}
\frac{V^s Q_{N-s}}{Q_N}=\prod_{i=1}^s\frac{V Q_{N-i}}{Q_{N-i+1}}\le 2^s
\end{equation}
and get the required upper bound.

As regards the term in square brackets in
(\ref{mayer_montroll}), instead, we expand the product and we get that
such a term is equal to
$$
1+\sum_{k=1}^{N-s}\!{N-s\choose k}\frac{1}{V^k}\!\int_\tV \!\!\dif^3\vett \tq^*_1
\ldots \int_\tV \!\!\dif^3 \vett \tq^*_k\prod_{i=1}^k \tilde{f}_s(\vett
\tq_1,\ldots, \vett \tq_s;\vett \tq^*_i) \tilde{F}^{(N-s)}_k(\vett
\tq^*_1,\ldots, \vett \tq^*_k)\ .
$$
We know an uniform upper bound for $\tilde{F}^{(N-s)}_k$, namely,
\begin{equation}\label{stima_bogoljubov}
\sup_{\vett \tq_1\in \tV,\ldots,\vett \tq_k\in \tV} \left|\tilde{F}^{(N-s)}_k
\right| \le \frac{(2e^{2b+1})^k} {1-\exp\left(2\rho e^{2b+1}I-1\right)}\ ,
\end{equation}
which holds in the hypotheses
$$
\frac{V+I}{2e^{2b+1}(\tilde{L}/L)(V(\tilde{L}/L)-I)}<1 \quad\mbox{and} \quad
\frac{V}{4e^{2b+2}(\tilde{L}/L)(V(\tilde{L}/L)-I)}<1\ .
$$
These conditions are certainly satisfied if hypothesis
(\ref{definizione_I}) holds.
Inequality (\ref{stima_bogoljubov}) has been proved\footnote{See the
  deduction of formula (4.10) in \cite{bogoljubov}.} by Bogolyubov et
al. in work \cite{bogoljubov}, which was dealing with pair potentials depending
  only on the distance between two particles. On the other hand, the
  hypothesis on the dependence on distance is not crucial, and a
  proof can be produced in the weaker hypothesis of potentials
  symmetric under the  exchange of $i$ and $j$. Indeed, the only
  difference from the proof given in \cite{bogoljubov} would be in the
  construction of the functions $\nu_i$, but we show in
  Lemma~\ref{lemma_simmetrizzazione} that it is possible to construct
  such functions in the present case, too. The $\nu_i$ are introduced
  in connection with the
  symmetrization operator $\pi_l$, which acts on the function $f(\vett
q_1,\ldots, \vett q_s)$ through
  the formula
$$
  \pi_l f(\vett q_1,\vett q_2,\ldots,\vett q_{l-1},\vett q_l,\vett q_{l+1},
  \ldots,\vett q_s)=f(\vett q_l,\vett q_2,\ldots,\vett q_{l-1},\vett
  q_1,\vett q_{l+1},\ldots,\vett q_s)\ .
$$
One has the following lemma, whose proof can be found in
Appendix~\ref{appendice_simmetrizzazione}.
\begin{lemma}\label{lemma_simmetrizzazione}
  Suppose there exists a positive constant $b$ such that, for all $s$
  and $(\vett q_1,\ldots,\vett q_s)$, the potential $\phi(\vett q_i,\vett
  q_j)= \phi(\vett q_j,\vett q_i)$ satisfies
\begin{equation}\label{stabilità_2}
  \sum_{1\le i<j\le s}\beta\,\phi(\vett q_i,\vett q_j)\ge -sb\ .
\end{equation}
Then, for all $s$, there exist measurable functions $\nu_i(\vett
q_1,\ldots, \vett q_s)$, having values in the interval $[0,1]$,
and such that
\begin{equation}\label{normalizzazione}
 \sum_{i=1}^s\nu_i(\vett q_1, \ldots, \vett q_s)=1\ , \quad \nu_k(\vett
 q_1,\ldots, \vett q_s)=\pi_k\nu_1(\vett q_1, \ldots, \vett q_s)\ ,
\end{equation}
with the inequality
\begin{equation}\label{consistenza}
\beta \sum_{j\neq i}\phi(\vett q_i,\vett q_j)>-2b
\end{equation}
holding if $\nu_i(\vett q_1,\ldots,\vett q_s)\neq 0$.
\end{lemma}

The only difference with respect to the functions used in \cite{bogoljubov}
is that  here the functions $\nu_i$ are
  not invariant under the rotation group. This, however, does not
  affect the proof, because it turns out that the proof given in
  \cite{bogoljubov} can be repeated word by word. This gives the upper bound
  (\ref{stima_bogoljubov}). 

For what concerns
$\tilde{f}_s$, we adapt to the present situation the reasoning followed
by O. Penrose in
work \cite{penrose}, which deals with hard--core potentials, in order
to provide  an upper bound in this case. Indeed, we notice that, by
condition (\ref{stabilità}), one has
\begin{equation}\label{maggiorazione_f_s}
1+\tilde{f}_s(\vett \tq_1,\ldots, \vett \tq_s;\vett y) =
\prod_{i=1}^s\exp\left[-\beta  \tilde{\Phi}(\vett \tq_i,\vett
  {\tilde{y}})\right] \le e^{2sb}\ .
\end{equation}
Defining
$$
g_i\= \exp\left[-\beta  \tilde{\Phi}(\vett \tq_i,\vett
  {\tilde{y}})\right] -1
$$
and $g_{i\pm}=\max(0,\pm g_i)\ge 0$,  we can use the upper bound in
(\ref{maggiorazione_f_s}) and the fact that $g_i\ge -1$, in order to prove by
induction on $s$ that
$$
1-\sum_{i=1}^sg_{i-}\le \prod_{i=1}^s(1+g_i)\le 1+e^{2sb} \sum_{i=1}^s
\frac{g_{i+}}{1+g_{i+}}\ . 
$$
We can use this relation in
(\ref{maggiorazione_f_s}) to obtain
$$
\left|\tilde{f}_s(\vett \tq_1,\ldots, \vett \tq_s;\vett y) \right| \le
\sum_{i=1}^s \max\left(g_{i-}, \frac{e^{2sb}\,g_{i+} }{1+g_{i+}}
\right)\ .
$$

Then, the integral over $\vett y$ of this quantity, for every choice
of $(\vett \tq_1, \ldots,\vett \tq_s)$, is smaller than $sI_s$, with
$I_s$ defined by (\ref{definizione_I_s}) if $s\ge 2$, while the case
of $s=1$ is trivial. If one, eventually, recalls that
$${N-s\choose k}<\frac{N^k}{k!}\ ,$$
one can then bound (\ref{mayer_montroll}) from above by
$$
\tilde{F}^{(N)}_s(\vett \tq_1,\ldots,\vett \tq_s)\le2^s \frac{\exp\left(2\rho
  e^{2b+1}sI_s\right) }{1-\exp\left(2\rho e^{2b+1}I-1\right)}\tilde{n}_s(\vett
\tq_1,\ldots,\vett \tq_s)\ .
$$

Observing that, for densities lower than
$1/4eI$, the denominator of the fraction in the r.h.s. is
larger  than $(\sqrt{e} -1)/\sqrt{e}$ , the thesis is finally proved
by going back to the initial
coordinates and using relation (\ref{F_tilde_N_s}). \begin{flushright}Q.E.D.\end{flushright}

\section{Proof of Lemma~\ref{lemma_simmetrizzazione}}\label{appendice_simmetrizzazione}

We show how to construct the functions $\nu_i$ having the properties
required in the lemma.

First, we fix a value for $s$ and consider the subsets
$A_i$ of the 
configuration space $X$, which are defined, for all $i\le s$, by
$$
A_i\=\left\{x\ :\ \beta\sum_{j\neq i} \phi(\vett
  q_i,\vett q_j)>-2b\right\}\ ,
$$
where $x\=(\vett q_1,\ldots,\vett q_s)$.
We observe now that, due to condition (\ref{stabilità_2}), one has
$$
\bigcup_i A_i=X\ .
$$
We can thus choose the function $\nu_1(x)$ as
$$
\nu_1(x)=N(x)\chi_{A_1}(x)\ ,
$$
where $\chi_A$ is the charachteristic function of the set $A$ and the function
$N(x)$, which is introduced to normalize the sum, takes the value $1/n$ if
$k$ belongs to $n$ sets $A_i$ but not to $n+1$ of them. Obviously,
$N(x)$ takes values in the set $\left\{1,1/2,\ldots,1/s\right\}$.

Then, if we construct $\nu_k$ by $\nu_k=\pi_k\nu_1$, it is
clear that the functions $\nu_i$ satisfy
condition (\ref{consistenza}), because 
$\pi_k\chi_{A_1}=\chi_{A_k}$ and because of the
definition of $A_i$. On the other hand, condition
(\ref{normalizzazione}) holds, too, because $\pi_k
N(x)=N(x)$. Indeed, the
belonging of $x$ to sets different from $A_1$ and $A_k$ is not
affected by the action of $\pi_k$, while $\pi_kx$ belongs to $A_1$ if
and only if $x$ belongs to $A_k$ and $\pi_kx$ belongs to $A_k$ if
and only if $x$ belongs to $A_1$. Thus, the number of sets to which $x$
belongs does not change under the action of $\pi_k$, and this
completes the proof.
\begin{flushright}Q.E.D.\end{flushright}

%
%

\end{document}